\documentclass[10pt, conference, letterpaper]{IEEEtran}
\IEEEoverridecommandlockouts
\usepackage[letterpaper, left=0.625in, right=0.625in, bottom=1.02in, top=0.73in]{geometry}
\usepackage{cite}
\usepackage{ctable}
\usepackage{lipsum}
\usepackage{xspace}
\usepackage{subcaption}
\usepackage[font=small]{caption}
\usepackage[cmex10]{amsmath}
\usepackage{mathtools,amssymb,lipsum}
\usepackage{lettrine}
\usepackage{bm}
\usepackage{algorithm,algorithmic}
\usepackage{environ}
\NewEnviron{NORMAL}{%
    \scalebox{2}{$\BODY$} 
}
\NewEnviron{HUGE}{%
    \scalebox{5}{$\BODY$} 
} 
\usepackage{array}
\usepackage{url}  
\usepackage{amsthm}
\usepackage{multirow}
\usepackage{color}
\usepackage{epstopdf}
\usepackage{endnotes}
\usepackage{framed}
\usepackage{cool}
\usepackage{enumerate}
\usepackage{url}
\usepackage[subnum]{cases}
\usepackage[nocomma]{optidef}
\usepackage{etoolbox}
\usepackage{amsmath,amssymb,amsfonts}
\usepackage{graphicx}
\usepackage{txfonts}
\usepackage[compact]{titlesec}         

\newtheorem{definition}{Definition}
\DeclareMathSizes{10}{9}{6}{5}
\usepackage{textcomp}
\usepackage{xcolor}

\DeclareMathOperator{\EX}{\mathbb{E}}

\def\BibTeX{{\rm B\kern-.05em{\sc i\kern-.025em b}\kern-.08em
    T\kern-.1667em\lower.7ex\hbox{E}\kern-.125emX}}
\begin{document}
\title{Semantic Enabled 6G LEO Satellite Communication for Earth Observation: A Resource-Constrained Network Optimization \\
}
\author{\IEEEauthorblockN{$^\dagger$$^*$Sheikh Salman Hassan, $^\dagger$$^*$Loc X. Nguyen, $^\ddagger$Yan Kyaw Tun, $^{\dagger\dagger}$Zhu Han, and $^\dagger$Choong Seon Hong}
\IEEEauthorblockA{$^\dagger$Department of Computer Science and Engineering, Kyung Hee University, Yongin, 17104, Republic of Korea\\
$^\ddagger$Department of Electronic Systems, Aalborg University, A . C. Meyers Vænge 15, 2450 København\\
$^{\dagger\dagger}$Department of Electrical and Computer Engineering, University of Houston, Houston, TX 77004-4005, USA}
Email: \{salman0335, xuanloc088, cshong\}@khu.ac.kr, ykt@es.aau.dk, zhan2@uh.edu.
\thanks{$^*$ Contribute Equally.} 
\thanks{This work has been accepted by the 2024 IEEE Global Communications Conference (GLOBECOM 2024), $\copyright$ 2024 IEEE. Copyright may be transferred without notice, after which this version may no longer be accessible. $\copyright$ 2024 IEEE. Personal use of this material is permitted. Permission from IEEE must be obtained for all other uses, in any current or future media, including reprinting/republishing this material for advertising or promotional purposes, creating new collective works, for resale or redistribution to servers or lists, or reusing any copyrighted component of this work in other works. *Dr. CS Hong is the corresponding author.}
}
\maketitle
\begin{abstract}
Earth observation satellites generate large amounts of real-time data for monitoring and managing time-critical events such as disaster relief missions. This presents a major challenge for satellite-to-ground communications operating under limited bandwidth capacities. This paper explores semantic communication (SC) as a potential alternative to traditional communication methods. The rationality for adopting SC is its inherent ability to reduce communication costs and make spectrum efficient for 6G non-terrestrial networks (6G-NTNs). We focus on the critical satellite imagery downlink communications latency optimization for Earth observation through SC techniques. We formulate the latency minimization problem with SC quality-of-service (SC-QoS) constraints and address this problem with a meta-heuristic discrete whale optimization algorithm (DWOA) and a one-to-one matching game. The proposed approach for captured image processing and transmission includes the integration of joint semantic and channel encoding to ensure downlink sum-rate optimization and latency minimization. Empirical results from experiments demonstrate the efficiency of the proposed framework for latency optimization while preserving high-quality data transmission when compared to baselines.
\end{abstract}
\section{Introduction}
Satellite-based Earth monitoring is critical for applications such as disaster management, particularly when terrestrial networks fail, and therefore it offers rapid and comprehensive coverage of disaster zones. High-resolution sensors enable land surveys and hazard management, generating vast imagery data. However, transmitting this data directly for analysis overwhelms current capabilities due to the sheer volume. For example, the Landsat system gathers extensive data requiring transmission, highlighting the need for innovative processing techniques \cite{landsat}. While complete data transmission offers advantages in change detection, it's inefficient for storage and transmission resources. To address this, upcoming research explores onboard processing and shifting data processing from ground stations to satellites \cite{on-board_processing}. This new workflow promises to improve downlink efficiency and reduce transmission resource requirements significantly.

Emerging as a key technology for ubiquitous 6G connectivity, non-terrestrial networks (NTNs) mainly composed of Low-Earth Orbit (LEO) satellite constellations hold promise in bridging the digital divide for remote and underserved regions \cite{LEO_comm}. However, the economic and efficiency-driven surge in LEO constellations necessitates further research to address the inherent limitations of these compact satellites. Specifically, their constrained onboard memory, communication bandwidth, and associated data transmission costs pose a significant challenge despite their economic advantages \cite{sat_chall}. To address this, semantic communication (SC) offers a promising solution by focusing on transmitting only critical information \cite{SC_swin}. By extracting the essence of the data and eliminating redundancies, SC significantly reduces the transmitted volume. Furthermore, SC employs a joint source-channel coding design, making it inherently more robust against environmental noise compared to traditional communication methods \cite{kurka2020deepjscc}. Integrating SC into satellite communication has the potential to alleviate bandwidth limitations and optimize data transmission. Therefore, this research proposes a solution to optimize data transmission and minimize latency, which is a critical parameter in time-sensitive communication for observational LEO satellites where SC goes beyond simple compression, focusing on understanding the data's meaning and transmitting only crucial elements. This significantly reduces transmission volume, frees up valuable resources, and ensures SC quality-of-service (SC-QoS). Additionally, SC's unique joint source-channel coding design enhances reliability by jointly optimizing the data and the transmission channel, leading to robust communication even in challenging environments. This work presents a novel approach for LEO satellites that minimizes semantic communication latency while guaranteeing SC-QoS requirements. To our knowledge, this is the first to jointly address these aspects in LEO satellite communications with the following key contributions:
\begin{itemize}
    \item We propose a novel architecture for a 6G communication network utilizing SC-enabled LEO satellites, which optimize resource utilization within constrained environments.
    \item Building upon this architecture, we formulate a critical latency minimization problem that ensures both SC and network QoS.
    \item To address this mixed-integer non-linear programming (MINLP) problem, we employ a combined approach. The meta-heuristic discrete whale optimization algorithm (DWOA) tackles integer variables, while a one-to-one matching game efficiently handles binary variables.
    \item The efficacy of our SC-enabled LEO satellite communication framework is validated through extensive simulations, demonstrating superior performance compared to the baselines across diverse network configurations.
\end{itemize}
\section{System Model}
\label{system_model}
\begin{figure*}[t]
    \centering
    \includegraphics[width=0.8\textwidth]{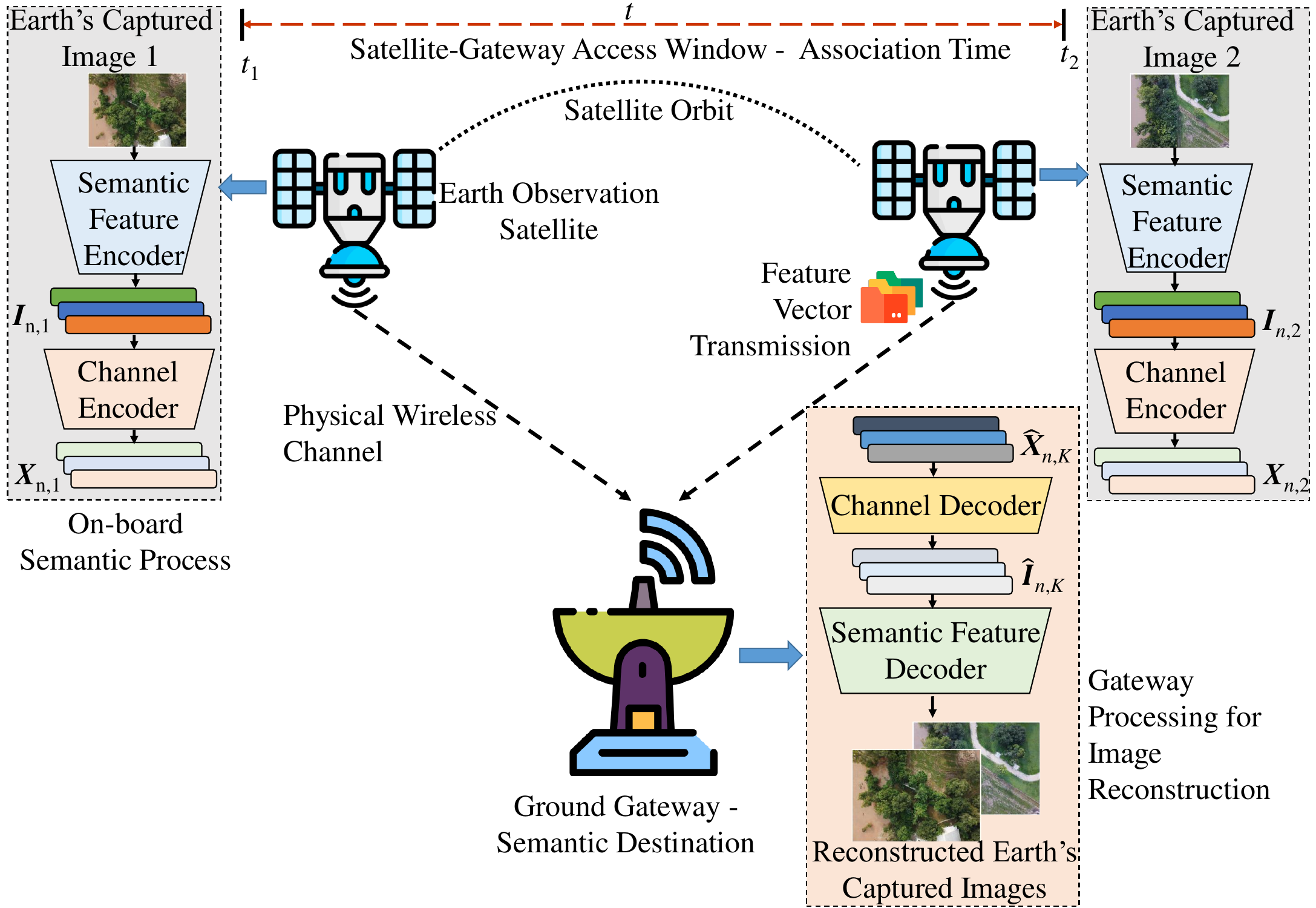}
    \caption{The architectural framework for facilitating multi-satellite semantic communication in Earth observation for disaster relief systems.}
    \label{system_model_fig}
\end{figure*}
\subsection{Network Model}
As illustrated in Fig. \ref{system_model_fig}, we assume a set $\mathcal{K}$ of $K$ LEO satellites and a set $\mathcal{I}$ of $I$ images captured by each satellite, all of which are positioned within the LEO constellation. These LEO satellites can establish a communication link with a ground terminal (GT), (i.e., gateway) during a specific temporal interval known as the access window, characterized by the shared visibility of both the satellite and the gateway, which is defined as $T_k$. Furthermore, each LEO satellite is equipped with a sensing camera designed to capture images of the Earth's surface and subsequently extract semantic features from these images through the utilization of an onboard artificial intelligence (AI) processor. Each satellite will transmit its computed semantic features during the access window above via an established communication link.

\subsection{Semantic Communication Framework}
Given the image $\boldsymbol{I}\in\boldsymbol{R}^{H\times W\times C}$ with indexing $i$ taken from the satellite camera, the semantic encoder will extract the main features $n$ of the image while eliminating the redundancy information. We denote the semantic encoder as $S_{\alpha}(\cdot)$ with the learning parameter $\alpha$. With the defined notation, the extracted features are represented as:
\begin{equation}
    f_{i}=S_{\alpha}(\boldsymbol{I}). \label{features}
\end{equation}
While the semantic encoder's role is feature selection, the channel encoder's responsibilities are compressing the extracted feature into a lower dimension and protecting the signal from the physical noise from the environment. The output of the channel encoder is denoted as follows:
\begin{equation}
    X_{i}=C_{\beta}(f_{i}), 
\end{equation}
where $C_{\beta}(\cdot)$ denotes the channel encoder with learnable parameter $\beta$. To eliminate or at least minimize the physical noise under various conditions, the proposed channel encoder uses the signal-to-noise ratio (SNR) feedback information from the receiver \cite{kurka2020deepjscc}. 

At the GT, the receiver will be equipped with two modules: channel decoder and semantic decoder. These modules will try to reverse the encoding process on the satellite. The following equation can denote the received signal over a wireless communication link:
\begin{equation}
    \hat{Y}_{i,k}= X_{i,k}H_{k}+N_{k},
\end{equation}
where $H_{I,k}$ is the channel between satellites $k$ and the GT, $N_{k}$ is the noise from physical environment. In this paper, we assume that the channel state information (CSI) is known at the GT (i.e., the receiver), and therefore, the signal can be transformed as follows: \cite{xie2021task}: 
\begin{equation}
    \hat{X}_{i,k}= (H^{H}H)^{-1}H^{H}\hat{Y}^{i,k}= X_{i,k}+ \hat{N}_{k},
\end{equation}
where $\hat{X}_{i,k}$ denotes the estimation of the encoded symbols of image $i$ at the LEO satellite (i.e., transmitter) $k$. The encoded symbols will be the input of the channel decoder to eliminate the noise and also decompress them to recover the extracted features of the original image:
\begin{equation}
    \hat{f}_{i,k}= C^{-1}_{\theta}(\hat{X}_{i,k}),
\end{equation}
where the $C_{\theta}^{-1}$ denotes the channel decoder with learning parameter $\theta$ and $\hat{f}_{i,k}$ is the estimated semantic features of the image $I$. Later, the image will be reconstructed by the semantic decoder using the above semantic features, i.e.,
\begin{equation}
    \hat{i}_{k}= S^{-1}_{\gamma}(\hat{f}_{i,k}),
\end{equation}
where $S^{-1}(\cdot)$ is the semantic decoder at the GT with the leaning parameter $\gamma$. To train the whole network in an end-to-end manner, the most common loss is the mean squared error (MSE) loss between the original and the reconstructed one: 
\begin{equation}
    d(i,\hat{i})= \textrm{MSE}(i,\hat{i}).
\end{equation}
To be more specific, this loss will calculate the value difference between each pixel in both images and force the network to reproduce the images that are identical to the original image. The quality of the reconstructed image is evaluated by the peak SNR (PSNR) metric as follows:
\begin{equation}
    \mathrm{PSNR}= 10\log_{10}\frac{\mathrm{MAX^2}}{\mathrm{MSE}}, 
\end{equation}
where $\mathrm{MAX}$ denotes the highest value of the image pixel per channel, which is determined $2^{n}-1$; for example, each color channel is presented by eight bits, and the value of $n$ is 8.   
\subsection{Semantic Feature Transmission Model}
We consider an orthogonal frequency-division multiple access (OFDMA) due to its flexibility in allocating limited network resources. This allows satellites to be assigned subcarriers based on their specific data requirements, efficiently handling uneven traffic patterns. Additionally, OFDMA inherently avoids inter-user interference within the spectrum. A set $\mathcal{U}$ of $U$ subcarriers is considered where a subset of these subcarriers will be assigned to each LEO satellite $k$. 

The downlink rate from the LEO satellite $k$ to the GT within subcarrier $u$ to transmit semantic feature information $\hat{f}_{i,k}$ as:
\begin{equation}
   R^u_k(\boldsymbol{\gamma}_k) = \sum_{u=1}^{U} \gamma_{k,u} B^u_k \log_2 \big( 1 + \Gamma_k \big), 
   \label{rate}
\end{equation}
where $B_k$ is the subcarrier bandwidth, $\boldsymbol{\gamma}_k = [\gamma_{k,1}, \cdots, \ \gamma_{k,U}]$ denotes the subcarrier assignment tuple for each LEO satellite $k$, which can be defined as:
\begin{equation}
    \gamma_{k,1} = \{0,1\}, \label{association_constraint_1}
\end{equation}
where $\gamma_{k,u}=1$ denotes that subcarrier $u$ is utilized by LEO satellite $k$, and $\gamma_{k,u}=0$ represents otherwise. We also provide the condition that only one subcarrier $u$ is assigned to each LEO satellite at each time slot to avoid network interference:
\begin{equation}
   \sum_{k=1}^{K} \gamma_{k,u} \leq 1. \label{association_constraint_2}
\end{equation}
Similarly, each LEO satellite $k$ can utilize at most a single subcarrier, which is given as follows:
\begin{equation}
    \sum_{u=1}^{U} \gamma_{k,u} \leq 1. \label{association_constraint_3}
\end{equation}  
Moreover, the SNR $\Gamma$ between LEO satellite $k$ and the GT can be defined as follows:
\begin{equation}
    \Gamma_k = G_k G_{\textrm{GT}} P_k \Big(\frac{c}{4\pi d_k f_c N_0} \Big)^2,
\end{equation}
where $G_u$ and $G_{\textrm{GT}}$ denote the wireless power antenna gains for the LEO satellite and GT respectively. $P_k$ is the LEO satellite transmit power, $c$ indicates the speed of light, $f_c$ denotes the carrier frequency, $d_k$ denotes the distance between the LEO satellite $k$ and the GT, and $N_0$ denotes the spectral density of the noise power to capture the interference generated by the LEO satellites in other constellations which utilize the same bandwidth spectrum and subcarrier $u$. 

Based on the semantic features obtained in (\ref{features}) and the achievable data rate calculated in (\ref{rate}), we can calculate the latency of transmitting the features from LEO satellite $k$ in subcarrier $u$ to the GT is:
\begin{equation}
    t(\boldsymbol{l_{X}}, \boldsymbol{\gamma}_k) = \frac{l_{X}}{R^u_k},
\end{equation}
where $l_{X}$ denotes the length of the transmitted data, and this length can be reduced or increased by adjusting the compression ratio. Extended signals provide receivers with additional information about channel noise, enhancing their ability to reconstruct the original signal accurately. Therefore, there will be a trade-off between the transmission latency and the quality of the reconstructed image.
\section{Latency Minimization Problem Formulation}
\label{prob_form}
Leveraging the aforementioned system model, we aim to minimize the average latency for transmitting semantic features from LEO satellites. However, this optimization must adhere to the network model's practical constraints. Specifically, sufficient semantic information must be transmitted within the access window to allow the GT to reconstruct the original image. Additionally, reliability and SC-QoS must be ensured by guaranteeing a minimum PSNR threshold that meets the desired image recovery quality. Under these assumptions, the average latency can be given as:
\begin{equation}
    \boldsymbol{O}(\boldsymbol{l_{X}}, \boldsymbol{\gamma}_k) = \frac{1}{K} \sum_{k \in \mathcal{K}} t(\boldsymbol{l_{X}}, \boldsymbol{\gamma}_k).
\end{equation}
The latency minimization problem can be formulated as:
\begin{mini!}|s|[2]                 
		{\substack{\boldsymbol{l_{X_I}}, \boldsymbol{\gamma}_k}} 
		{\boldsymbol{O}(\boldsymbol{l_{X_I}}, \boldsymbol{\gamma}_k)  }{\label{P1}}{\textbf{P1:}}	
		\addConstraint{\gamma_{k,1} = \{0,1\}, \quad \forall k \in \mathcal{K}, ~\forall u \in \mathcal{U} }, {\label{prob_association_constraint_1} }
		\addConstraint{\sum_{u=1}^{U} \gamma_{k,u} \leq 1, \quad \forall k \in \mathcal{K}}, {\label{prob_association_constraint_2}}
		\addConstraint{ \sum_{k=1}^{K} \gamma_{k,u} \leq 1, \quad \forall u \in \mathcal{U}}, {\label{prob_association_constraint_3}}
		\addConstraint{\EX(\mathrm{PSNR}_k(\boldsymbol{X}))} \geq \Psi_k, \quad \forall k \in \mathcal{K}, {\label{PNSR_Constraint_4}}
  	\addConstraint{0 \leq \gamma_{k} t_k \leq T_k, \quad       \forall k \in \mathcal{K}},      {\label{Access_window_Constraint_5}}
\end{mini!}
where constraints (\ref{prob_association_constraint_1}), (\ref{prob_association_constraint_2}), and (\ref{prob_association_constraint_3}) guarantee that the ground station will assign only a single subcarrier to each LEO satellite, and every subcarrier could only be utilized by a single LEO satellite at each time slot for semantic features transmission. Constraint (\ref{PNSR_Constraint_4}) ensures that the semantic features should be reliable and meet the SC-QoS threshold from each LEO satellite $k$. Constraint (\ref{Access_window_Constraint_5}) ensures the communication efficiency of LEO satellites within the access window, i.e., the coverage period of the GT.
It can be observed that the formulated problem is MINLP which is difficult to solve due to NP-hard. To address this problem, we provide a solution approach in the next section.
\section{Proposed Algorithm}
\label{sol_algo}
To address the formulated MINLP, we utilize the block coordinated descent (BCD) approach, where we decomposed the main problem in (\ref{P1}) into two subproblems according to the nature of decision variables and then solved iteratively.
\subsection{Image Reliable Retrieval (SC-QoS) Problem}
In this subproblem, we minimize the total transmission latency while ensuring the quality of the received images by determining the transmitted length of the data. Given the feasible value $\boldsymbol{\gamma_k}$ of subcarrier assignment, the problem can be decomposed as follows:
\begin{mini!}|s|[2]                 
		{\substack{\boldsymbol{l_{X}}}} 
		{\boldsymbol{O}(\boldsymbol{l_{X}}, \boldsymbol{\gamma_k} )}    {\label{P1.1}}{\textbf{P1.1:}}	
        \addConstraint{ (\ref{PNSR_Constraint_4})~\textrm{and}~(\ref{Access_window_Constraint_5}).
        }     
\end{mini!}
The compression ratio affects not only the transmission latency but also the quality of the reconstructed images. As shown in (\ref{PNSR_Constraint_4}), the GS demands different service quality from each satellite, which results in different transmitted lengths for each satellite. To tackle the challenge of integer variables in problem \textbf{P1.1}, we leverage the meta-heuristic discrete whale optimization algorithm (DWOA) \cite{DWOA}. DWOA adapts the operators of the standard WOA to work effectively with integer decision variables. This process involves initializing a population of candidate integer solutions within the feasible search space. Subsequently, DWOA iteratively refines these solutions through exploration and exploitation phases, which are specifically adapted to handle discrete values. The exploitation can be expressed by the following equations:
\begin{equation}
    \Vec{l}_{X} (\tau+1)=\left \{ \begin{array}{ll}{\Vec{l}_{X}^{*}(\tau)-\Vec{A}\cdot\Vec{D},} & {\textrm{p $<$0.5,}} \\ 
      \\{\Vec{D}' \cdot e^{bm} \cdot cos (2\pi m) + \Vec{l}_{X}^{*}(\tau),} & {\textrm{p $\geq$0.5,}}\end{array}\right.
\end{equation}
where ${\Vec{l}_{X}^{*}}(\tau)$ denotes the best solution at the current iteration $\tau$. \textit{p} is a random variable, which is used to decide between two hunting behaviors of whales. On the other hand, the searching agents explore other solutions by updating their location toward a random agent in the population and can be expressed as:
\begin{equation}
    \Vec{l}_{X}(\tau+1)=  \Vec{l}_{X_{rand}}(\tau)-\Vec{A}\cdot\Vec{D}.  \label{sub4}
\end{equation}
Each candidate's fitness is evaluated based on the objective function. In addition, to deal with the constraints, we integrate them into the objective function as a punishment value if violated. The algorithm then iteratively updates solution positions and selects the best solution found so far, continuing until a pre-defined stopping criterion is met. This approach allows DWOA to search for high-quality solutions to problems with integer constraints efficiently, which is described in Algorithm \ref{alg:DWOA}.
\begin{algorithm}[t]
	\caption{\strut DWOA for SC-QoS Problem} 
	\label{alg:DWOA}
	\begin{algorithmic}[1]
	\STATE{\textbf{Input:} $\mathcal{N}$: number of searching agents, $\mathcal{CRS}$: compression ratio set, $\mathcal{K}$, $\Psi_k$ and the access window $T_{k}$, $\gamma_{k}$, the maximum number of iterations \textit{MaxIT}.}
 	\STATE{\textbf{Output:} The optimal transmitted length $\boldsymbol{l_{X}^{*}}$ and the communication latency.}
        \STATE{Initialize the whale population, $\boldsymbol{l_{X}}$ = [$l_{n1}, x_{n2}, ..., x_{nk}$] $\in \mathcal{CRS}$  $(n=1,2,...,\mathcal{N})$. Calculate the communication latency of each agent and determine the optimal decision $l^{*}_{X}$, $\tau \gets 0$.}
        \WHILE{$\tau \leq MaxIT $}
            \FOR{$n \gets 1 $ to $\mathcal{N}$ (each searching agent) }
            \STATE {Update vector \textit{a, A, C, m} and \textit{p}.}
            \IF{$p < 0.5$}
                \IF{$ |A| < 1$}
                    \STATE{Update the $\boldsymbol{l_{X}}$ by the first sub-equation of (18).}
                \ELSE
                    \STATE Update the  transmitted length by (\ref{sub4}).
                \ENDIF
            \ELSIF{$p\geq 0.5$}
            \STATE Update the $\boldsymbol{l_{X}}$ by the second sub-equation of (18).
            \ENDIF
            \ENDFOR
            \STATE Calculate the transmission latency of each agent and update $\boldsymbol{l_{X}^{*}}$ if there is a better transmission length.
            \STATE $\tau \gets \tau +1$. 
        \ENDWHILE
	\STATE{\textbf{Output:} Optimal transmitted length $\boldsymbol{l_{X}^{*}}$.}
	\end{algorithmic}
\end{algorithm}
\subsection{Subcarrier Allocation to LEO Satellite}
This subproblem optimizes the binary optimization part of the \textbf{P1} to address the subcarrier allocation to each LEO satellite $u$ from GT. Similarly, by having the optimal PSNR solution $\boldsymbol{X_I}^*$ from the subproblem in (\ref{P1.1}), we can decompose the subcarrier assignment problem as follows:
\begin{mini!}|s|[2]                 
		{\substack{\boldsymbol{\gamma}_k}} 
		{\boldsymbol{O}(\boldsymbol{l_{X}^*}, \boldsymbol{\gamma_k} )  }{\label{P1.2}}{\textbf{P1.2:}}	
	\addConstraint{ 
        (\ref{prob_association_constraint_1}),
        ~(\ref{prob_association_constraint_2}),
        ~(\ref{prob_association_constraint_3}), ~\textrm{and}
        ~(\ref{Access_window_Constraint_5}),
        } 
\end{mini!}
To solve this binary optimization problem, we utilize a matching game for the subcarrier allocation to the LEO satellite. According to constraints (\ref{prob_association_constraint_1}) and (\ref{prob_association_constraint_2}) in problem \textbf{P1.2}, each subcarrier can be allocated to at most one LEO satellite, and each LEO satellite can only have one subcarrier allocated to it. Therefore, we reformulate the subcarrier allocation problem as a two-sided one-to-one matching game. We first define a one-to-one matching game for the subcarrier allocation to the LEO satellite at the GT.
\vspace{-0.08in}
\begin{definition}
Consider two disjoint sets of players, $\mathcal{K}$ and $\mathcal{U}$. The two-sided one-to-one matching game $\varsigma$ : $\mathcal{K}$ $\rightarrow \mathcal{U}$ for the subcarrier allocation is defined as:
\begin{enumerate}
    \item $\varsigma(u) \subseteq \mathcal{K} ~and ~|\varsigma(u)| \in \{0,1\}, \quad \forall u \in \mathcal{U}$;
    \item $\varsigma(k) \subseteq \mathcal{U} ~and ~|\varsigma(k)| \in \{0,1\}, \quad \forall k \in \mathcal{K}$;
    \item $ k= \varsigma(u) \Leftrightarrow u = \varsigma(k), \quad \forall u \in \mathcal{U}, ~\forall k \in \mathcal{K}.   $
    \vspace{-0.05in}
\end{enumerate}
\end{definition}
\noindent Here, $|\varsigma(.)|$ is the cardinality of the matching outcome $\varsigma(.)$, where the outcome of the matching is the allocation mapping between a set $\mathcal{K}$ of LEO satellites and $\mathcal{U}$ of subcarriers. Furthermore, conditions (1) and (2) in the definition guarantee that one subcarrier can be allocated to at most one satellite at a time and that one satellite can only have one subcarrier allocated to it. Finally, condition (3) ensures that if LEO satellite $k$ is matched with subcarrier $u$ then subcarrier $u$ must also be matched with LEO satellite $k$.
\begin{algorithm}[t]
	\caption{\strut One-to-One Matching for subcarrier Allocation} 
	\label{alg:association}
	\begin{algorithmic}[1]
	\STATE{\textbf{Initialization:} $\mathcal{K}, ~\mathcal{U}, ~\boldsymbol{P}, ~\mathcal{K}^{\textrm{un}}$=$\mathcal{K}, ~\mathcal{U}_k$=$\mathcal{U}, \forall k \in \mathcal{K}$, a set of LEO satellites requested to subcarrier $u$, $\mathcal{K}^{u,\mathrm{req}}= \emptyset$, and a set of rejected LEO satellites from subcarrier $u$, $\mathcal{K}^{u,\mathrm{rej}}, \forall u \in \mathcal{U}$;   }
        \STATE{LEO satellite $k$ designs the preference list $\succ_k$ with (\ref{leo_pref});    }
        \WHILE{$\sum_{k\in \mathcal{K}} \sum_{u\in \mathcal{U}} q_{ku} \neq 0 $ }
        \FOR{$ k = 1$ to $|\mathcal{K}^{\textrm{un}}|$}
        \STATE{Find $u = \underset{u\in \mathcal{U}}{\mathrm{argmax}} ~\zeta_k(u) $}.
        \STATE{Make a request to the GT by setting $q_{ku}$ = 1.}
        \ENDFOR   
        \FOR{$ u = 1$ to $U$}
        \STATE{Update $\mathcal{K}^{u,\mathrm{req}} \leftarrow \{k : q_{ku} = 1, \forall k \in \mathcal{K} \}$  }.
        \STATE{Construct the preference of GT for its available subcarrier according to (\ref{GT_pref}). }
        \STATE{Find $ k = \underset{k\in \mathcal{K}}{\mathrm{argmax}} ~\zeta_u(k) $}.
        \STATE{Allocate subcarrier $u$ to LEO satellite $k$.}
        \STATE{Update $\mathcal{K}^{u,\mathrm{rej}} \leftarrow \{\mathcal{K}^{u,\mathrm{req}} \setminus k$\} }.
        \STATE{Update $\mathcal{U}_k \leftarrow \{\mathcal{U}_k \setminus u \}  \forall k \in  \mathcal{K}^{u,\mathrm{rej}}$ }.
        \ENDFOR 
        \STATE{Update $\mathcal{K}^{\textrm{un}} \leftarrow \mathcal{K}^{\textrm{un}} \cap \{\mathcal{K}^{1,\mathrm{rej}} \cup \cdots \cup \mathcal{K}^{U,\mathrm{rej}}  \}$ }.
        \ENDWHILE
	\STATE{\textbf{Output:} Optimal subcarrier allocation $\boldsymbol{\gamma}^*$}.
	\end{algorithmic}
\end{algorithm}

\textbf{Preference of players:} Matching is carried out based on
the preference profiles constructed by the LEO satellites and the GT
over subcarriers to prioritize potential pairings based on their
information. To define the matching game, let us denote $\succ_k$
and $\succ_u$ the preference profiles of LEO satellite $k \in \mathcal{K}$ and subcarrier $u \in \mathcal{U}$. Furthermore, let $\zeta_k(u)$ and $\zeta_u(k)$ be the preference functions of LEO satellite $k \in \mathcal{K}$ for subcarrier $u$ and subcarrier $u \in \mathcal{U}$ for LEO satellite $k$, respectively. Note that the preference function of each satellite must satisfy the constraint (\ref{Access_window_Constraint_5}), therefore the preference function of LEO satellite $k$ for subcarrier $u$ is given by:
\begin{equation}
    \zeta_k(u) = \varrho_k(u) R_k^u( \boldsymbol{\gamma}_k),  \label{leo_pref}
\end{equation}
where $\varrho_k(u)$ is an indicator function with the value of $1$ if constraint (\ref{Access_window_Constraint_5}) for LEO satellite $k$ is satisfied and $0$ otherwise. Each LEO satellite prefers to be allocated to the subcarrier that provides the highest communication rate. For instance, $u \succ_k u'$ implies that LEO satellite $k$ prefers subcarrier $u$ over subcarrier $u'$ to be allocated, i.e., $\zeta_k(u) > \zeta_k(u')$. Similarly, we can define the preference function of subcarrier $u$ for LEO satellite $k$ as:
\begin{equation}
     \zeta_u(k) = R^u_k( \boldsymbol{\gamma}_k).   \label{GT_pref}
\end{equation}
This preference reflects that the GT desires to match each subcarrier $u$ to a LEO satellite $k$ that achieves the maximum communication
rate on that subcarrier to maximize the total communication
rate which ultimately minimizes the latency, which is the objective function of \textbf{P1}.
\begin{definition}
A stable matching $\varsigma^*$ is achieved if there is no blocking pair $(k, u)$, where a pair $(k, u)$ is a blocking pair when $u \notin \varsigma(u), k \notin \varsigma(k)$, and  $u \succ_k \varsigma(u)$ and $k \succ_u \varsigma(k)$.
\end{definition}
Since the proposed game is implemented exactly like the standard deferred acceptance algorithm \cite{roth2008deferred}, it guarantees a stable matching. The output of the algorithm is the optimal subcarrier assignment $\boldsymbol{\gamma}^*$. The pseudocode of the proposed one-to-one matching game-based subcarrier assignment algorithm is shown in Algorithm \ref{alg:association}. The overall solution approach is summarized in Algorithm \ref{alg:sequence}.

\begin{algorithm}[t]
	\caption{\strut The sequence of overall Proposed Scheme} 
	\label{alg:sequence}
	\begin{algorithmic}
	\STATE{\textbf{Step 1:} Given the feasible value $\boldsymbol{\gamma_k}$ of subcarrier assignment, we run DWOA to determine the length of transmitted signals.}
 	\STATE{\textbf{Step 2:} With the temporary transmitted lengths from DWOA, we optimize the subcarrier for each satellite by the One-to-One Matching from Algorithm~\ref{alg:association}.}
	\STATE{\textbf{Step 3:} Finally, we re-run the DWOA to determine the final transmitted lengths for each satellite given the optimal subcarrier.}
	\end{algorithmic}
\end{algorithm}
\setlength{\arrayrulewidth}{0.10mm}
\setlength{\tabcolsep}{1pt}
\renewcommand{\arraystretch}{0.7}
\begin{table}[t]
\centering
\renewcommand{\arraystretch}{0.8}
\caption{Simulation Parameters}
\label{sim_tab}
\scalebox{1.1}{
\begin{tabular}{|c||c||c|}
\hline
    \textbf{Notation}& \textbf{Definition}& \textbf{Value} \\ \hline \hline
    $f_c$ & Carrier Frequency & $[20-30]$~GHz\\ \hline
    $B_u$ & Bandwidth & $500$~MHz\\ \hline
    $G_u~$ & LEO Satellite Antenna Gain & $33.13$~dBi \\ \hline
    $T_k~$ & Access Window & $60$~Seconds\\ \hline
    $d~$ & LEO Satellite Altitude & $786$~km \\ \hline
    $N_0~$ & Noise Power & $-43$~dB\\ \hline
    $G_{\textrm{GT}}~$ & Ground Terminal Antenna Gain & $~34.2$~dBi \\ \hline
    $p_u~$ & LEO satellite Transmit Power & $~10$~W\\ \hline
    $T_{\textrm{orb}}$ & Orbital Period & $100$~min \\ \hline 
    $\mathcal{CRS}$ & Compression Ratio Set & $[\frac{4}{128} ; \frac{5}{128} ... \frac{11}{128}; \frac{12}{128} ]$ \\ \hline
\end{tabular}}
\end{table}
\section{Simulation Results Evaluation}
\label{simul}
To validate the proposed joint DWOA and matching game scheme for SC in GS-LEO networks, we conduct simulations and results analyses. We establish the simulation environment and introduce benchmark schemes for comparison. Subsequently, we evaluate the performance of the proposed scheme. The main parameters are summarized in Table \ref{sim_tab}. To evaluate the effectiveness of SC-enabled GS-LEO communication, we compare our proposed scheme against the following benchmarks:
\begin{itemize}
    \item \textit{OnlyWhale}: In this scheme, the GS solely employs the DWOA to optimize the transmit length for guaranteeing the PSNR requirements. While the resource block is randomly allocated to each satellite.
    \item \textit{MatchingOnly}: Here, the GS utilizes only a matching scheme to allocate each resource block for the satellites, while a greedy algorithm randomly searches the transmission length of the data. 
    \item \textit{Random}: This baseline randomly assigns the resource blocks to the satellites and the transmitted length is achieved by the greedy algorithm. 
\end{itemize}
\begin{figure}[t]
    \centering
    \includegraphics[width=0.75\columnwidth, height=2in]{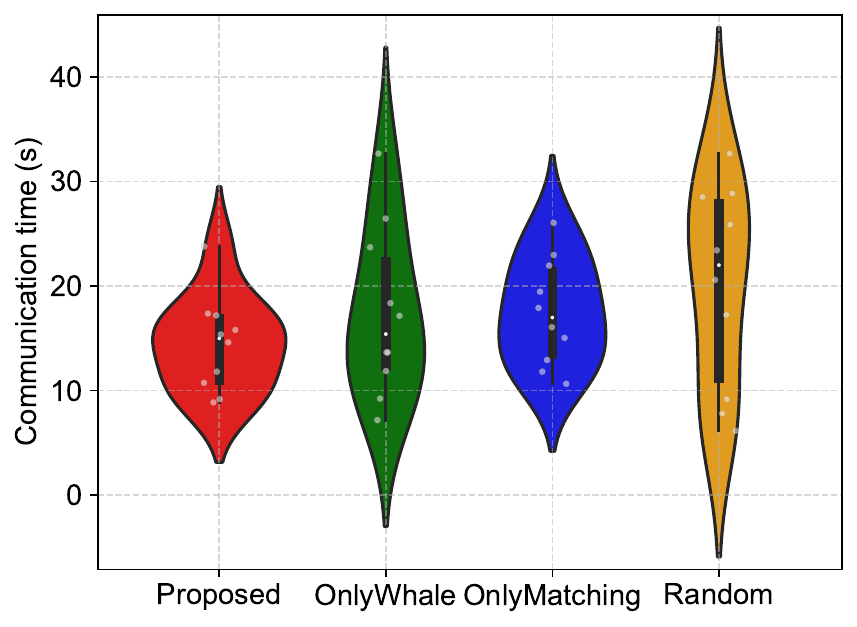}
    \caption{Communication performance evaluation.}
    \label{Communication_Perfromance}
\end{figure}
\begin{figure}[t]
    \centering
    \includegraphics[width=0.75\columnwidth]{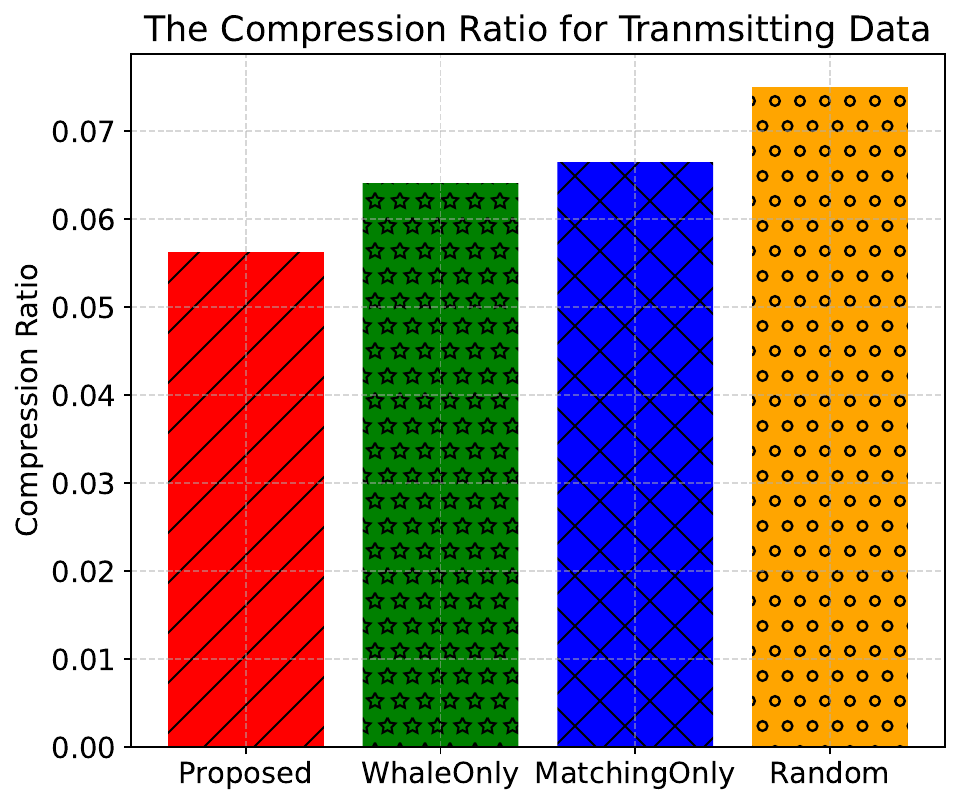}
    \caption{Compression ratio comparison.}                 
    \label{Compression_Impact}
\end{figure}

\begin{figure*}[t]
    \centering
    \begin{subfigure}[t]{0.245\textwidth}
        \centering
        \includegraphics[width=\textwidth]{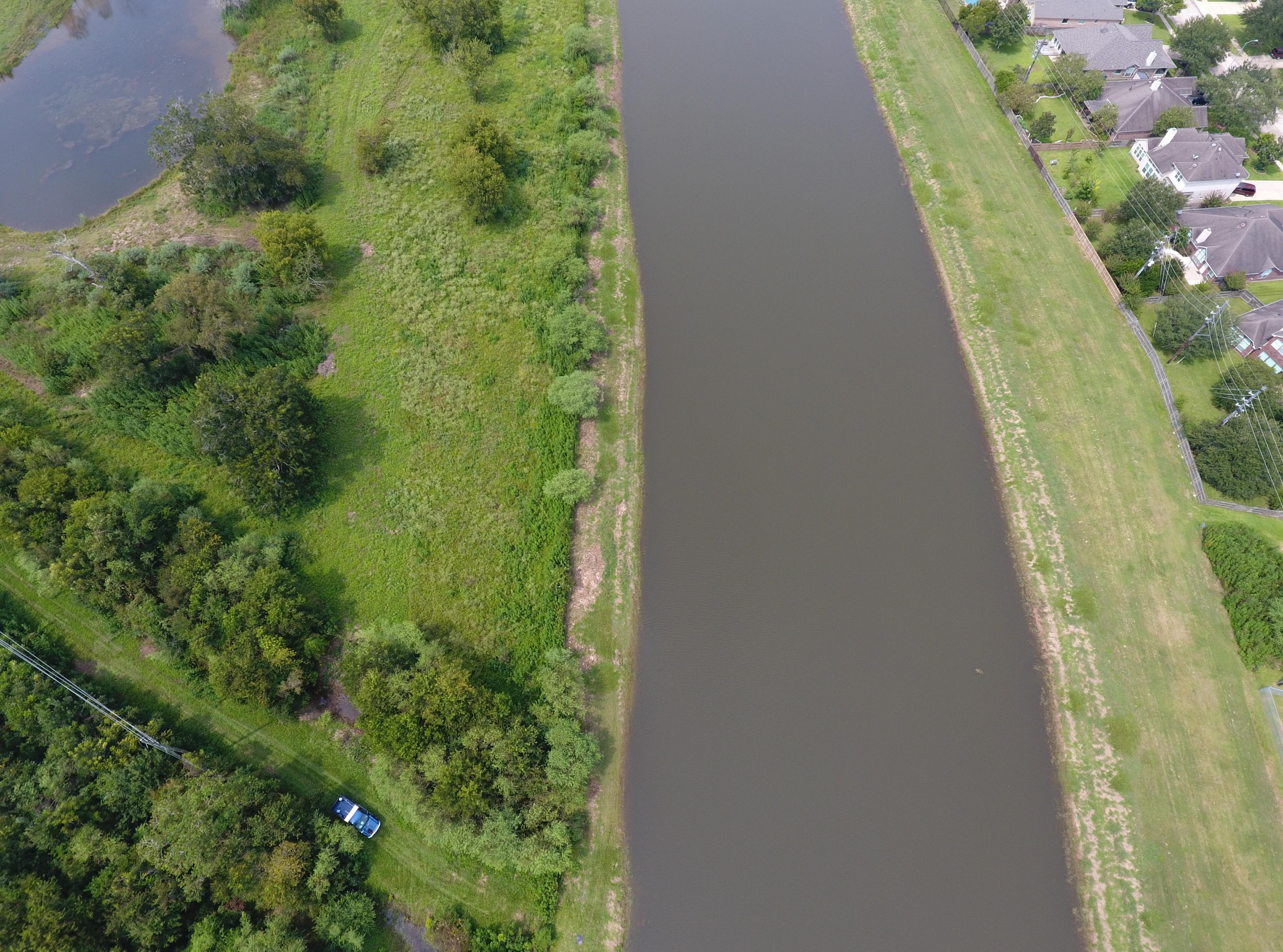}
        \caption{Original Image}
        \label{Original}
    \end{subfigure}
    \hfill
    \begin{subfigure}[t]{0.245\textwidth}
        \centering
        \includegraphics[width=\textwidth]{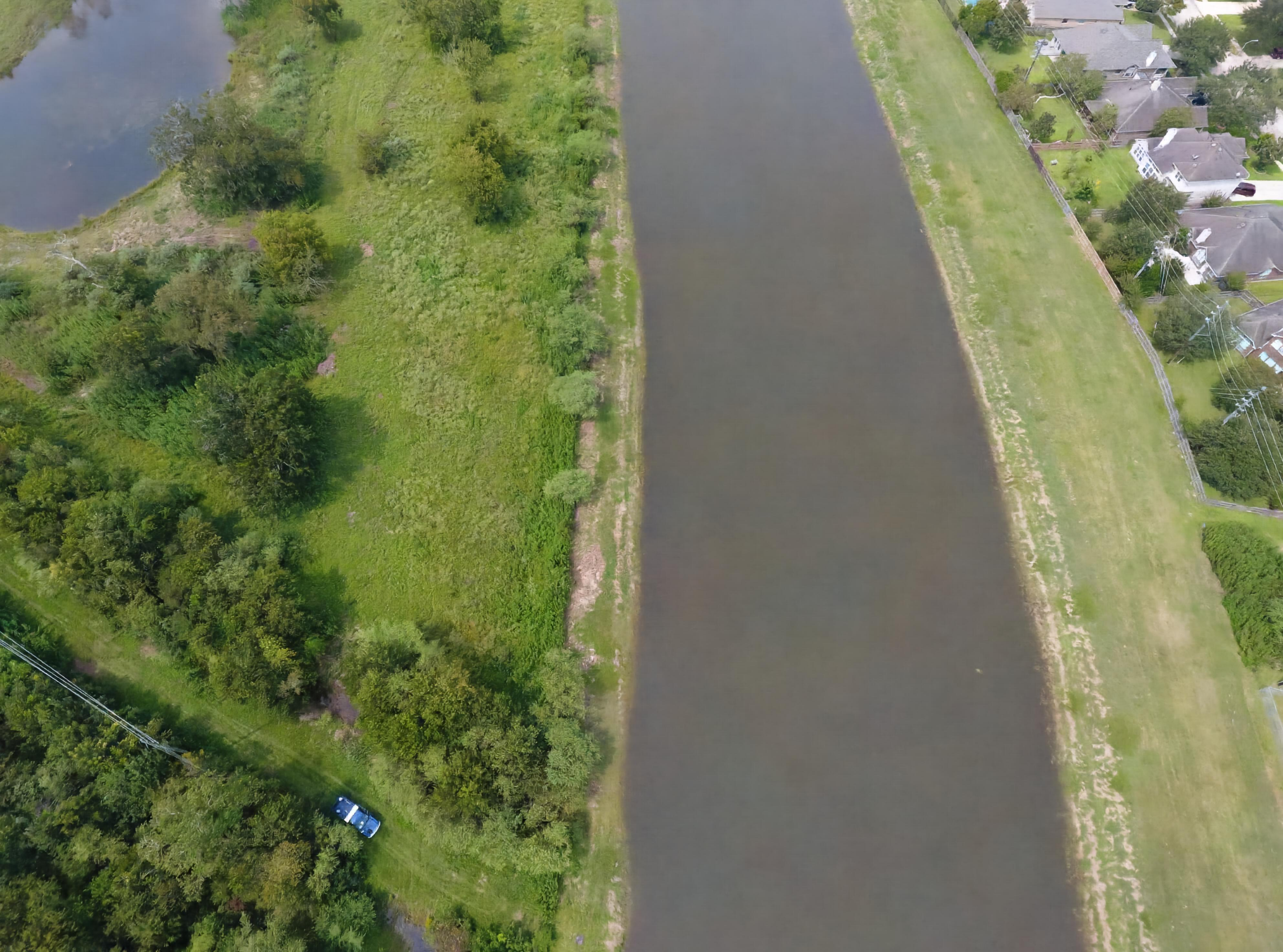}
        \caption{Compression Ratio: 4/128}
        \label{48CR}
    \end{subfigure}
    \hfill
    \begin{subfigure}[t]{0.245\textwidth}
        \centering
        \includegraphics[width=\textwidth]{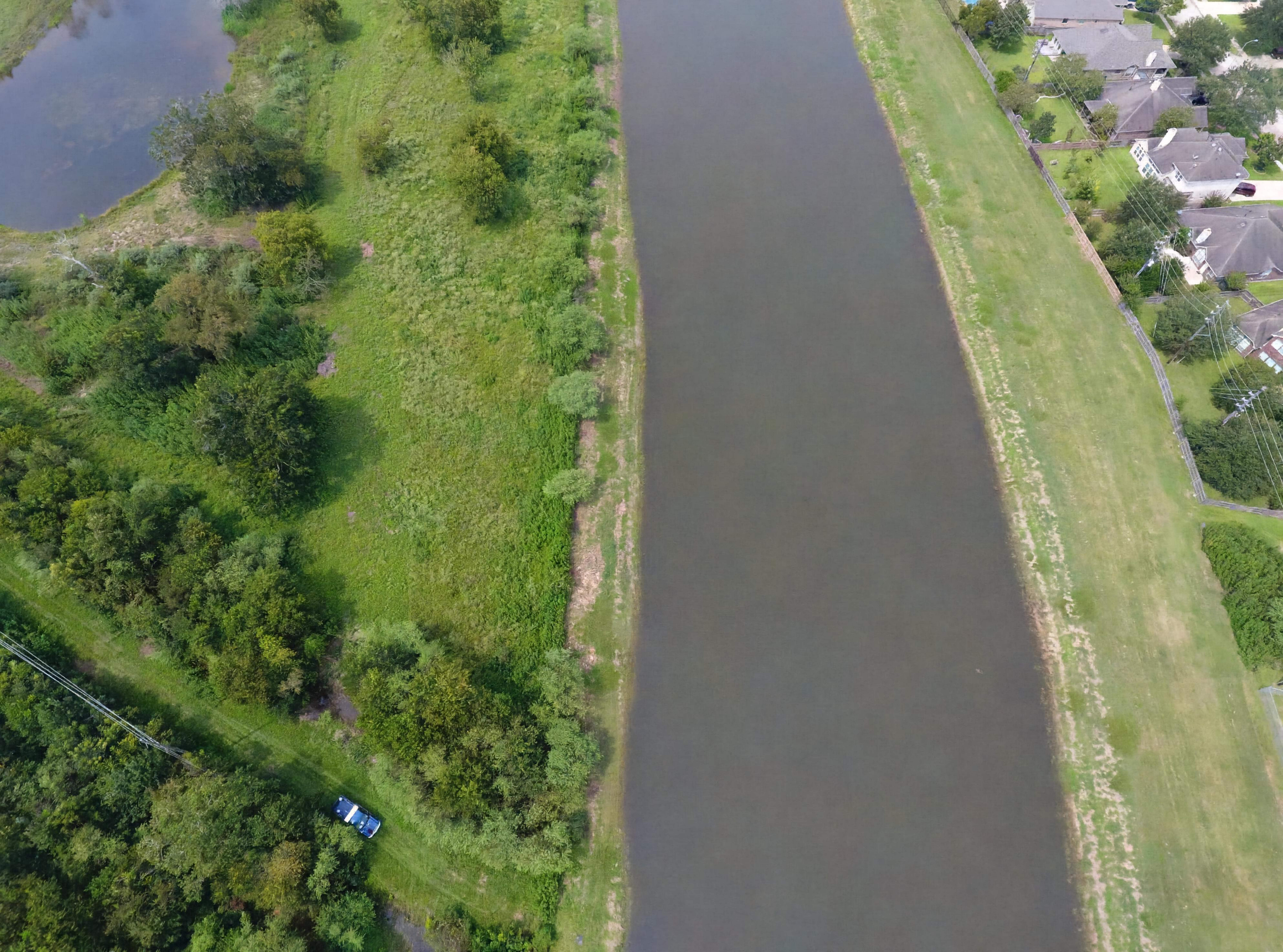}
        \caption{Compression Ratio: 8/128}
        \label{96CR}
    \end{subfigure}
    \hfill
    \begin{subfigure}[t]{0.245\textwidth}
        \centering
        \includegraphics[width=\textwidth]{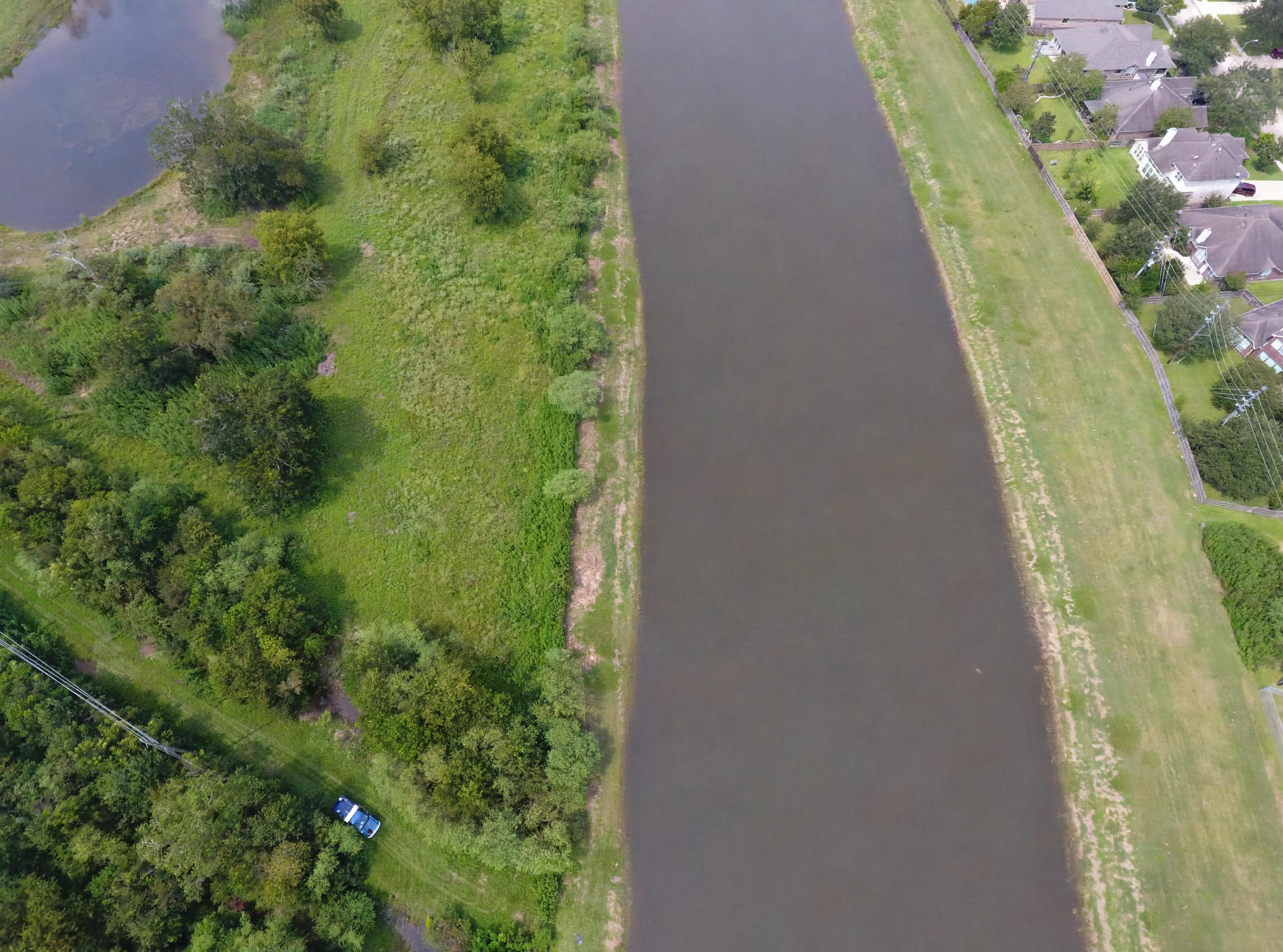}
        \caption{Compression Ratio: 12/128}
        \label{144CR}
    \end{subfigure}
    \caption{The original image and the reconstruction images with different compression ratios under the SNR = 3dB.}
     \label{Reconstructedimage}
\end{figure*}
\begin{figure}[t]
    \centering
    \includegraphics[width=0.85\columnwidth, height=2in]{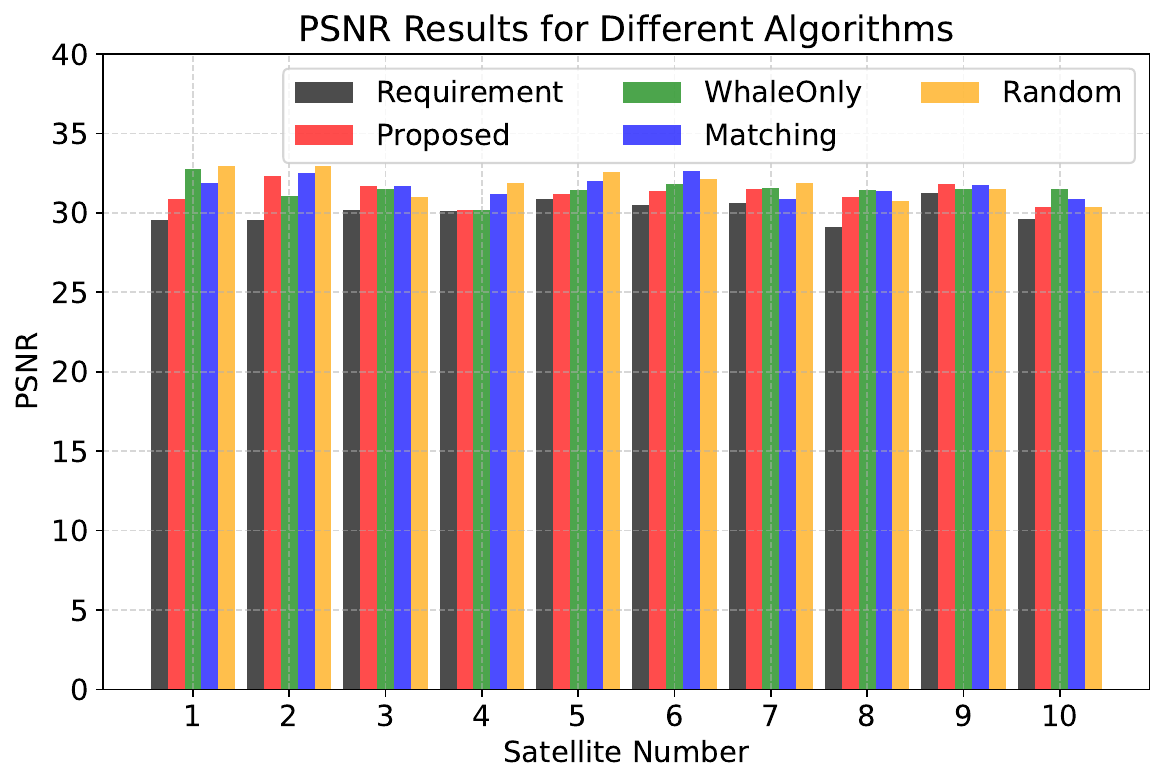}
    \caption{PSNR evaluation.}
    \label{PSNR_evaluation}
\end{figure}

Fig. \ref{Communication_Perfromance} depicts the communication latency comparison of the proposed joint DWOA and matching game scheme against three baseline approaches. The objective is to minimize latency while ensuring both SC and network QoS requirements. As evident from the figure, the proposed algorithm achieves superior performance in terms of communication latency for the required transmission. In simpler terms, the proposed scheme transmits data with lower latency compared to the baselines while still meeting the necessary SC and network QoS constraints.

Fig. \ref{Compression_Impact} shows the average compression ratio of the satellites for all the scenarios. The proposed scheme achieves the lowest compression ratio, which equivalently indicates the lowest transmit length while satisfying the requirement of QoS for each satellite. Compared to the baseline schemes, the proposed scheme exhibits a significant advantage. The \textit{WhaleOnly} scheme achieves a relatively large compression ratio performance due to insufficient resource allocation. Therefore, it suffers from higher communication latency as shown in Fig. \ref{Communication_Perfromance}. The \textit{MatchingOnly} and \textit{Random} perform poorly in the problem of determining the transmitted length, which leads to a high compression ratio. In conclusion, the proposed scheme offers a superior trade-off between image quality and communication efficiency.

Fig. \ref{Reconstructedimage} presents original and reconstructed images from the FloodNet dataset \cite{9460988}, which offers high-resolution unmanned aerial system (UAS) imagery with comprehensive semantic annotations detailing post-disaster damage. This showcases the generalizability of our semantic model to satellite datasets, as evidenced by its effectiveness on UAS imagery. It can be observed that the semantic communication model has achieved an outstanding visual result for different compression ratios, which makes it difficult to spot the difference by the human eyes. However, with the PSNR metric, there is a significant difference in quality for those compressions. To be specific, the PSNR and MS-SSIM values are [27.23, 0.84], [28.42, 0.89], [29.34, 0.91] corresponding to the compression ratios: 4/128, 8/128, 12/128, respectively. These results have effectively demonstrated the effect of transmitting lengths on the performance of the receiver. The longer the transmitted signals are transmitted, the better the receiver performs in capturing the channel noise and constructing more quality images.

Fig. \ref{PSNR_evaluation} further validates the effectiveness of our proposed algorithm by assessing its performance in achieving PSNR requirements for individual satellites under network dynamics. Unlike some baseline algorithms that exhibit inconsistent performance across satellites, the proposed scheme demonstrates consistent results. It successfully meets or comes very close to the required PSNR for most satellites in various network conditions. This consistency highlights the robustness of our approach in handling diverse PSNR demands across the network while ensuring quality. Additionally, the proposed scheme guarantees that the PSNR threshold is never violated.
\section{Conclusion}
\label{conclusion}
This work presented a novel approach for Earth observational data transmission using an SC-enabled LEO satellite network for GS communication. The proposed framework leverages multiple LEO satellites with onboard processing capabilities. These satellites extract critical semantic features from captured images and transmit them to the GS, enabling reconstruction of the original image. This strategy achieves efficient spectrum utilization and significantly reduces network latency. Consequently, it overcomes the limitations of traditional LEO satellite communication for Earth observation applications, where high latency is particularly detrimental due to the large data volumes involved. Furthermore, extensive simulations demonstrate the superior performance of the proposed algorithm compared to baseline approaches, validating its effectiveness. 
\ifCLASSOPTIONcaptionsoff
\newpage
\fi
\bibliographystyle{IEEEtran}
\bibliography{main}

\begin{thebibliography}{10}
\providecommand{\url}[1]{#1}
\csname url@samestyle\endcsname
\providecommand{\newblock}{\relax}
\providecommand{\bibinfo}[2]{#2}
\providecommand{\BIBentrySTDinterwordspacing}{\spaceskip=0pt\relax}
\providecommand{\BIBentryALTinterwordstretchfactor}{4}
\providecommand{\BIBentryALTinterwordspacing}{\spaceskip=\fontdimen2\font plus
\BIBentryALTinterwordstretchfactor\fontdimen3\font minus
  \fontdimen4\font\relax}
\providecommand{\BIBforeignlanguage}[2]{{%
\expandafter\ifx\csname l@#1\endcsname\relax
\typeout{** WARNING: IEEEtran.bst: No hyphenation pattern has been}%
\typeout{** loaded for the language `#1'. Using the pattern for}%
\typeout{** the default language instead.}%
\else
\language=\csname l@#1\endcsname
\fi
#2}}
\providecommand{\BIBdecl}{\relax}
\BIBdecl

\bibitem{landsat}
``U.{S}. geological survey: What are the landsat collection 1 level-1 data
  product file sizes?''
  \url{https://www.usgs.gov/faqs/what-are-landsat-collection-1-level-1-data-product-file-sizes},
  2022.

\bibitem{on-board_processing}
V.-P. Bui, T.~Q. Dinh, I.~Leyva-Mayorga, S.~R. Pandey, E.~Lagunas, and
  P.~Popovski, ``On-board change detection for resource-efficient earth
  observation with {LEO} satellites,'' in \emph{Proc. of the IEEE Global
  Communications Conference}, Kuala Lumpur, Malaysia, Dec. 2023.

\bibitem{LEO_comm}
S.~S. Hassan, Y.~M. Park, Y.~K. Tun, W.~Saad, Z.~Han, and C.~S. Hong,
  ``Space{RIS}: {LEO} satellite coverage maximization in {6G} sub-{TH}z
  networks by {MAPPO} {DRL} and whale optimization,'' \emph{IEEE Journal on
  Selected Areas in Communications}, Feb. 2024, early access.

\bibitem{sat_chall}
R.~Xie, Q.~Tang, Q.~Wang, X.~Liu, F.~R. Yu, and T.~Huang,
  ``Satellite-terrestrial integrated edge computing networks: {A}rchitecture,
  challenges, and open issues,'' \emph{IEEE Network}, vol.~34, no.~3, pp.
  224--231, Mar. 2020.

\bibitem{SC_swin}
L.~X. Nguyen, Y.~L. Tun, Y.~K. Tun, M.~N.~H. Nguyen, C.~Zhang, Z.~Han, and
  C.~S. Hong, ``Swin transformer-based dynamic semantic communication for
  multi-user with different computing capacity,'' \emph{IEEE Transactions on
  Vehicular Technology}, Feb. 2024, early access.

\bibitem{kurka2020deepjscc}
D.~B. Kurka and D.~G{\"u}nd{\"u}z, ``Deepjscc-f: Deep joint source-channel
  coding of images with feedback,'' \emph{IEEE Journal on Selected Areas in
  Information Theory}, vol.~1, no.~1, pp. 178--193, Apr. 2020.

\bibitem{xie2021task}
H.~Xie, Z.~Qin, and G.~Y. Li, ``Task-oriented multi-user semantic
  communications for vqa,'' \emph{IEEE Wireless Communications Letters},
  vol.~11, no.~3, pp. 553--557, Dec. 2021.

\bibitem{DWOA}
Y.~Li, Y.~He, X.~Liu, X.~Guo, and Z.~Li, ``A novel discrete whale optimization
  algorithm for solving knapsack problems,'' \emph{Applied Intelligence},
  vol.~50, no.~10, pp. 3350--3366, Jun. 2020.

\bibitem{roth2008deferred}
A.~E. Roth, ``Deferred acceptance algorithms: {H}istory, theory, practice, and
  open questions,'' \emph{International Journal of Game Theory}, vol.~36,
  no.~3, pp. 537--569, Jan. 2008.

\bibitem{9460988}
M.~Rahnemoonfar, T.~Chowdhury, A.~Sarkar, D.~Varshney, M.~Yari, and R.~R.
  Murphy, ``Floodnet: A high resolution aerial imagery dataset for post flood
  scene understanding,'' \emph{IEEE Access}, vol.~9, pp. 89\,644--89\,654, Jun.
  2021.

\end{thebibliography}
\end{document}